# A Desktop-Centric Design Space for Direct Object Examination and Visualization in Mixed-Reality Environments

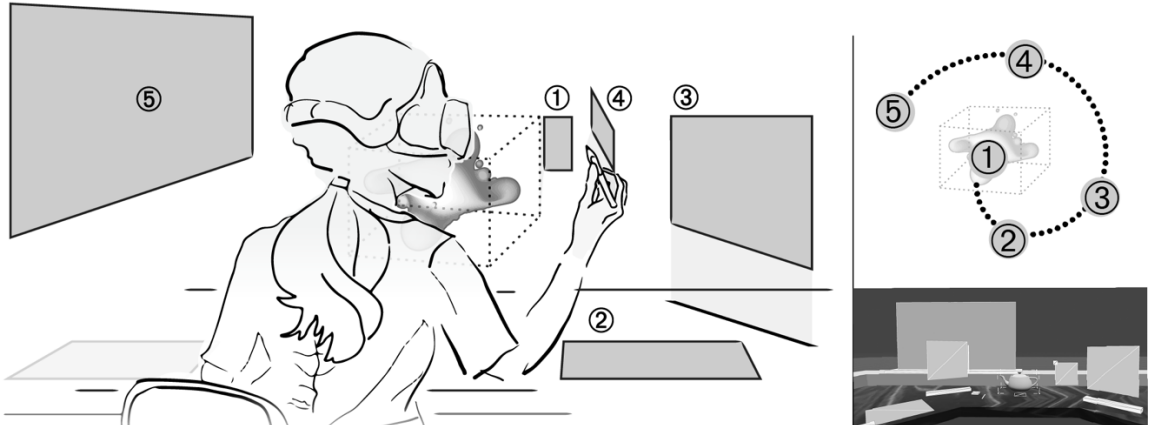

**Figure 1:** (left) Illustration depicting user interacting in MR with affordances designed around five zones (1) object zone, (2) desktop zone, (3) free-floating panel zone, (4) hand/pen zone, and (5) wall zone; (top right) numbering scheme for zones centered around main object being inspected; (bottom right) screenshot of working protype of interaction zones

Sam Johnson-Lacoss*
Carleton College, johnsonlacosss@carleton.edu

Santiago V. Lombeyda
California Institute of Technology, slombey@caltech.edu

S. George Djorgovski
California Institute of Technology, djorgovski@caltech.edu

*Mixed reality (MR) environments are bound to become ubiquitous as MR technology becomes lighter, higher resolution, more affordable, and overall becomes a seamless extension of our current work and living spaces. For research scientists and clinicians focused on understanding 3D phenomena or patient pathologies within the context of the larger human anatomy, that means a necessary evolution of their workstations currently only utilizing 2D interfaces for everyday communication, logistics and data analysis. MR technologies bring forth immersive 3D representations coexisting in our natural spaces, while allowing for richer interconnected information displays, where 3D representations greatly aid in the detailed understanding of physical structures, spatial relationships, and 3D contextualization of 2D measurements, projections, abstractions, and other data details. We present a breakdown of the different interaction zones and modalities into a design space that best accommodates the creation of applications for users engaged through MR technologies in precise object-centric data analysis within the ergonomic confines of their desktop physical spaces ([Figure 1](#)).*

**ACM Reference Format:**

## 1 Introduction

Measuring, comparing, even navigating complex 3D structures all require cognitive harmony between 3D structure understanding and the deeper 2D abstraction of patterns. It is thus the natural inclination to turn to Virtual Reality (VR) and Mixed Reality (MR) spaces to create richer and more informative 3D immersive experiences. Ailments arising from conventional VR technologies, such as cybersickness, lack of a sense of presence, loss of navigation context from traveling around virtual spaces, and even difficulty in selection of objects beyond natural reach (mostly using ray-base techniques) are resolved through the use of tabletop or desktop VR/MR where a user's interaction mostly happens with objects at arms-reach on a desktop (both virtual and real) while on a sitting position [2]. Meanwhile any remaining issues arising from the isolation and disconnectedness with a user's natural environment when engaging with completely immersive VR spaces, are resolved by turning to MR, where 3D objects (and 2D screens placed into 3D space) are digitally displayed into their real space giving users back a sense of understanding of the surroundings. So, for MR application developers the goal is usually to place these virtual objects into our real space as seamlessly as possible, in order to create natural and fluid experiences.

## 2 Background

Currently, VR/MR tools designed for visualization and data analytics mostly focus on fully immersive experiences. Most of these tools use a combination of space anchored affordances such as 3D buttons, levers, or signage that must be accessed via direct touch or casting ray from a hand held controller, or controller based input such as buttons, triggers, joystick, or pads, or sometimes though virtual buttons or other affordances attached to the controller virtual representation. Alternatively, 3D first person games often use displays mounted or attached to the viewers camera or screen (heads-up-displays or HUDs). These choices for both VR applications as well as games are the result of the application and user interface designer's lack of control of a user's location *by design*.

As our goal is to create more natural and fluid interfaces for scientists and clinicians who do most of their data inspection and analytical work while at a desk (sitting or standing in place), we on the other hand can assume a de facto desktop or tabletop to weave our interfaces onto. This exact methodology has proven to be useful in general visualization endeavors [3] were the data is a directly manipulable object in front of the user, hovering right above their desktop, or inspecting volumetric radiology [6], as well as in more niche clinical tools like tumor selection inside human lungs [4].

Research, such as [1], has also has measured the ways in which layouts of information inside VR/MR arranged close proximity to the user (in semi-circular formation) impact their ability to perform tasks and minimize mental effort and frustration, and maximize visuo-spatialmemory. This same study participants completed various tasks relating to data visualization in both 2D and 3D involving five major zones: *the wall*, *the floor*, *the tabletop*, *the body*, and *the cockpit* (free-standing and movable representations in close proximity to the user); where the tabletop, the floor, and the cockpit were designed for manipulative visualization, while the other for observation at a distance.

Complimentary, research [5] has compared different input modalities, such as hand based and gesture based versus controller and pointer based versus hybrid virtual wrist-attached input devices (virtual smartwatch), only to reveal that while some input modalities, such as hand based, can be overall relatively more comfortable, the choice for better input device is ultimately task dependent, hardware precession dependent, and of course, dependent on all design and implementation choices.

Furthermore, tangible markers utilized in MR environments for data visualization and analysis, have proved overall to improve time and accuracy of tasks [7]. We are specifically interested in pen utilization, as humans are capable of very precise hand actions, which can be further refined, as exemplified by millenia of examples from artists and artisans for all kinds of endeavors. Pen based has proven to show promise in VR/MR environment, with hardware having been around for more than 40 years, such as the Polhemus 6DOF stylus (3D pen used in direct object manipulation in VR documents as early as [12] and later as part of the Responsive Workbench [13]). Pen centric research continues to grow, with richer types of interaction explored [10], and 3D positional space scanners such as UltraLeap actively supporting tool-tracking [11]. And with the use of pen tools in MR is still a growing field, its effectiveness has only just started to be evaluated. This includes [8] showing the increased effectiveness of different modalities of pen selection tools for instance, all while new hardware being actively developed, such as the VRSketchPen [9].

However, tangible interfaces have been explored for data visualization and analytics, both across different tactile spaces (including the creation of objects for different interactions [16]), to direct actions on tabletops in general ([15]), to more specific contexts, such as for visualization analytics ([17]). And while for specific analytical tasks, tangibles would be very hard to generalize into a design space, for general tasks and within the context of a given exploratory narrative, they can prove valuable.

Additionally distance based models, proxemics, create a methodology to scrutinize the locations as a factor in the interaction and resulting ability to better understand visualization [14]. Results show a preference for locked scaling, for using the space to arrange data logically, for having precise control in the selection processes (in the form of mouse in their study), and for the overall comfort level for a natural space. All are shown as keys to developing a fluid space for visualization and analytics.

We adopt all this knowledge, and embrace the constraints of a desktop/tabletop interaction spaces to tasks and case studies centered around understanding of 3D visualization and direct study and analysis of 3D objects, to formulate the clusters of affordances or taxonomy of interaction zoned into a design space for MR, which we empirically validated through paper prototyping, and user centric design alongside stakeholders, i.e. scientists and clinicians.

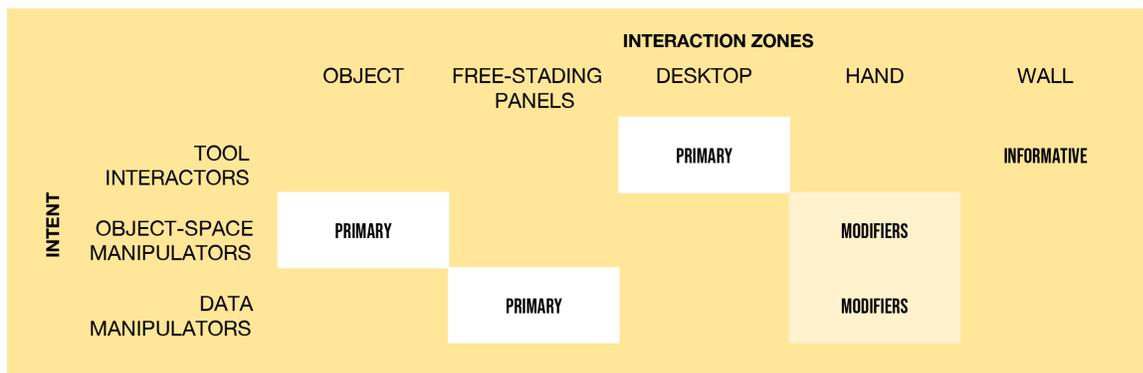

**Figure 2:** Grid showing the different taxonomies used to define the design space, and the projected use of each

# 3  Approach

Our approach to defining a design space that covers the basic interactions for data analysis and visualization within a physical desktop space utilizing MR technologies, begins with user-centric methodology, ie. direct collaboration with scientists and clinicians, interviews, and careful analysis of their current work methodologies. It quickly became clear as our first user constraint, that scientists and clinicians come pre-set mental models on how they understand their data. These are classically two dimensional, and even as part of their common work routine they are unlikely to project this 2D data into 3D, even if/when it was a possibility. The measurement of a tumor is easier on a 2D projection than in 3D where you need to find a larger context. Note as an extension of this, when deeper analysis was done programmatically or through utilization of machine-learning methodologies, they found their results easier to understand in 2D, though 3D context was often helpful.

While 3D data can embody extremely valuable information, but in most contexts designed to enable deeper cognitive understanding it is usually secondary to 2D data. Humans are used to operating efficiently both 2D and 3D data in a real-world environment. Meanwhile the ability to present 2D screens in immersive 3D environments with enough resolution (pixel number and density) and visual stability (smooth head tracking and high frame rate) is a relatively recent development.

Our approach is bifurcated by our understanding of possible interactions in the MR space into intent versus location. Further detail of the approach will be discussed in the details of the proposed taxonomy and design space to follow (Figure 2).

# 4  Classification by Intent

Digital mixed mediums combining 2D and 3D information are already common to many professional endeavors. They range from 3D development platforms like Unity or Epic Games' Unreal Engine, to 3D modelers like Blender or Autodesk's Maya, to 3D visualization systems like KitWare's ParaView. All these, along with software designed to systematically arrange a wealth of tools and interactions modalities, such as Adobe Photoshop, can serve as inspiration and as models on how professionals interact with data across 2D and 3D, mixed and side by side representations.

Due to the goal of visualization for data inspection and analysis, we can categorize actions with the scene into Tool Interactors, Object Space Manipulators, and Data Space Manipulators.

## 4.1 Tool Interactors

General interaction meant to affect the overall state of the application falls under the umbrella of tool interactors. As such, their location may be best suited to expected placements and anchored to the virtual/physical space, but relatively separate from the main object being studied.

## 4.2 Object Space Manipulators

Object Space Manipulators (OSM) constitute actions meant to affect only the superficial representation of the object being studied. These are also changes that recontextualize the data's presentation such as selecting a subset of data, highlighting, or expansion of multi-component packages. Because these manipulators are intimately attached to the visual representation, their location may be best suited to always being close to the object being studied, and even anchored to a corner of the bound box or major direction of the bounding sphere.

## 4.3 Data Manipulators

Conversely, Data Manipulators (DM) refer to interactions that modify the data itself, not just its representation. This would constitute such actions as simulating blood flow through a vessel being explored in the space in order to monitor blood pressure, which the model would potentially reflect in turn. As filters and generators that may radically affect the representation of the object being studied, but ultimately represent a transformation not separate or independent execution pathway, the location of these may be best suited in the vicinity of the object being studied, with possibly some of the same visual qualities on the object being studied (such as ability to move around the space, as well as initial elevation)..

# 5 Classification by Zone

We define our desktop-centric MR design space as a set of zones of interaction mostly within arm's reach of the researcher. Each offers specific modalities and affordances aimed for different modes of interactions in service of the inspection process, data manipulation, and overall tool state, all ultimately with the larger goal of enabling a natural and fluid process for understanding of the phenomena, its structures or pathologies.

We began with examination of the methodologies needed to support the inspection processes that scientists or clinicians would partake in, and set out to achieve an efficient and intuitive division of the environment given a set of precepts:

- First, an environment must *minimize clutter* as much as possible without losing any information that would be communicated. This means that the user should be never overwhelmed nor underwhelmed by the amount of data with which they are presented and should be able to navigate data without needing to trudge through anything
- This in turn plays into intuition: the environment must have *as little a learning curve as possible* in order to allow for users from any background to work in the environment with minimal confusion
- Finally, to make a space both *efficient and intuitive*, the environment must have *clear* divisions as to *where each of its functions should lie* so that the user of the space may interact with it in a logical and ergonomic way. A user must be able to quickly conceptualize where they should search for a certain type of representation or interaction.

Working with these constraints, we created the division of five zones upon which interaction with the space may occur or data may be presented.

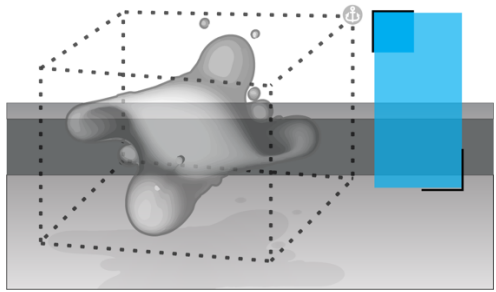

**Figure 3:** The Object Zone: Illustration of UI elements being attached to object itself

## 5.1 The Object Zone

Due to their proximity to the 3D object, any information in this zone adds clutter to the object area, muddying the representation. Therefore, components mounted to the 3D object should be the main position for OSMs due to the zone's proximity to the three-dimensional representation, and any information displayed in this zone should be hyper-specific to the 3D representation, not to the overarching data (Figure 3). This would allow for more focused manipulation, and it would prevent the object's immediate vicinity from being too crowded.

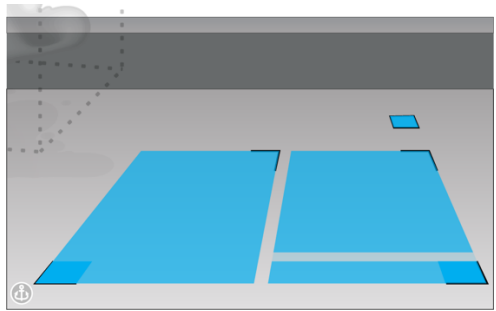

**Figure 4:** **The Desktop Zone: Illustration of UI elements being to accessible locations of the tabletop in front and to the side of the user**

## 5.2 The Desktop Zone

The desktop is too far from either the 2D or 3D representations for its zone to be used for effective manipulation of data, yet it is in quite close proximity to the user themself. Moreover, the desktop is the only zone apart from the hand whose virtual representation is identical to that of real space. As such, the desktop constitutes a resting place, upon which the user can confidently place their hands when not interacting with any data. Interactions on the desktop should thus be ones that relate to tool interactors, i.e. the user's experience in the space such as generalized settings, saving and loading spaces, and if any OSMs or DMs relate directly to the user's hands, the ability to swap these manipulators must be duplicated to the desk to allow for ergonomic switching when not focused on the data (Figure 4).

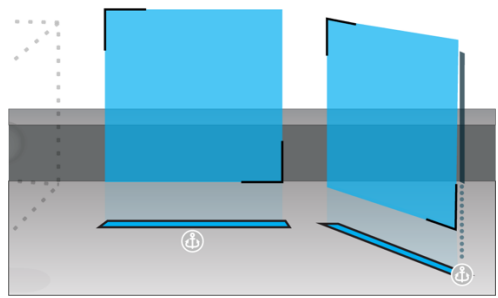

**Figure 5:** **The Free-Standing Panels Zone: Illustration of UI elements and panels hovering above desktop, with affordances to allow movement of them and visual hints of relative position to the desktop**

## 5.3 Free-Standing Panels Zone

Free-standing panels should visualize data that, although linked closely to the object's representation, is not necessarily directly connected to its visualization, as panels of this zone stand freely in the space similarly to a computer monitor yet may also be held and repositioned like paper or a tablet, all of which are common ways to visualize 2D data in real space. The presence of 2D data on free-standing panels thus implies that some 2D panels merit the classification of "object" themselves, as 2D and 3D representations are both parallel foci of the space. As such, zone 2 should contain subclass manipulators and information hyper-specific to the representations shown instances of this zone (Figure 5). This sub-zone therefore extends the Object zone, meaning that components mounted to data-bearing free-standing panels should contain information that is hyper-specific to the two-dimensional representation, and they should be the hub for DMs associated with the representation.

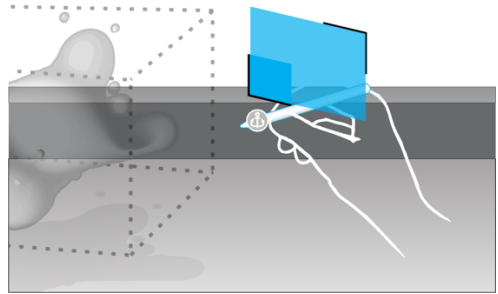

**Figure 6:** Hand+Pen Zone: Illustration of UI elements attached to tip of interaction tool

### 5.4 The Hand (and The Pen) Zone

Due to their proximity to the main interactor of the space, any component mounted to the hand must closely tie into interaction. This could constitute informational panels relating to the user's currently-selected tool, but too much information mounted to the hand would lead to a more claustrophobic space overall, as the addition of more information to this zone sacrifices the articulate qualities of the hand. As such, attachment of information to the hand should be kept to the absolute minimum, and anything mounted to the hand must be easily dismounted so as to remove feelings of distress brought on by attachment to the hand without physical reinforcement (Figure 6).

This was well-accomplished through tools like [4], as the information mounted to the hand was only a 3D model of the currently-equipped tool which instantly visually relayed the function of said tool. It also utilized a controller-based interaction system, adding the physical feedback of holding the tool in real space, a luxury this system does not have. However, this problem may be circumvented through a hybrid model of both hand- and controller-based interaction. In order to preserve the ergonomics of the current hand-based space, this controller must be accessible to all users, it must be easy to pick up from within the virtual space, and it must have pinpoint accuracy for tools that require precision. From previous unreleased work, we had experience with virtual reality environments that utilized a Leap Motion to track a cylindrical object that acts as a paint brush, to simulate the tracking of a pen with a button in space. The pen is small and thus easy to pick up and place, it is one of the most intuitive implements to use in a virtual environment due to its real-life prevalence, and its simple cylindrical shape lends itself well to non-cluttered representations of the currently-equipped tool. Furthermore, the ease of placement allows for the efficient equipping and unequipping of tools. As such, the hand zone becomes the location upon which data about the current representation of the pen's tool resides.

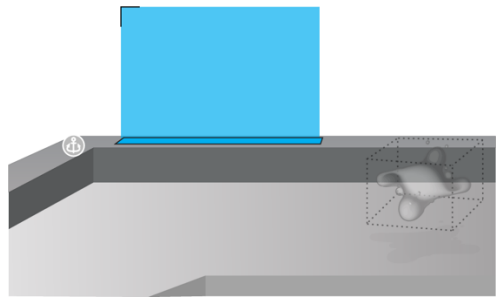

**Figure 7:** The Wall Zone: Illustration of panels anchored to back wall, most likely outside the reach of the user

### 5.5 The Wall Zone

The zone on the wall is the furthest from the user's experience both spatially and in interactive potential. Since it is so far from the user, the user is unable to efficiently change what it displays, and it becomes difficult to focus on the area. This zone should therefore be entirely redundant in its information depiction. As it is not in focus during visualization, and as it cannot be changed, it must show static information about both the virtual and real environments such as notifications about the space, time since last save, operating system notifications, and a view of the current desktop (Figure 7). This allows for effective communication of non-data-specific information while still maintaining relevance to the space.

# 6 Design Decisions

Note as a comparison, [1] defined five interaction zones, where participants consistently rated combinations of the wall, tabletop, and cockpit displays most highly. Due to the design of their environment, wall displays are much more similar to free-standing displays at a medium distance from the user than they are to that of the wall/ambiance zone in our space, while the cockpit resembles our free-standing panels at a close proximity. The results of their study thus support the style of interaction that we have proposed with respect to the desktop and the free-standing zones, as the placement of data manipulators and two-dimensional visualizations among free-standing panels provides egocentric and grounded locations for manipulation and visualization similar to that of the wall and cockpit simultaneously, while the tabletop acts as a grounding force, that which participants noted allows for more specific concentration on the task at hand.

## 6.1 Validating Design Decisions through Paper Prototyping

Having devised this subdivision for zones of interaction, we prototyped potential layouts for the division of these tools in real space through both sketches and three-dimensional paper prototypes (paper correlates to virtual objects and panels).

## 6.2 Validating Design Decisions through General Case Study

There are many common threads between case studies used for scientific research and those for clinicians. While each data very much specifies the way the data is best represented, and the goals for each researcher determines the overall process and use of the exploratory tool, we can still surmise main stages of interaction through a common narrative. The common framework we have established is grounded on this general case study. As the user enters the space, it shall become clear how following the guidelines of this design space, makes use of each of the zones in a manner conducive to increased efficiency, and demonstrating the validity of the summaries, taxonomies, and overall design decisions previously described. The subsections below demonstrate each step in the case study's narrative, along with supplementary images and videos taken from a working prototype of the space. At the foot of the description of each step is a table of active interaction zones and their use status (Figure 8, 9, 10, 11, 12, 13).

### 6.2.1 NARRATIVE STEP 01: Entering the Environment.

The user opens the software to their desktop environment. A Desk panel is open with "Load/Export/…" options to load the space, while a Free-standing panel (collapsed) indicates that there are outstanding notifications concerning the space. On the Wall is a panel that mirrors the desktop and the Free-standing panel's notifications.

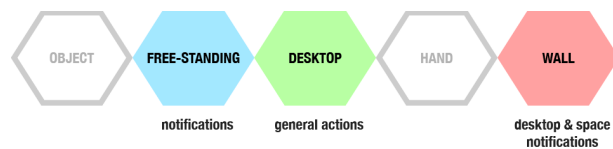

Figure 8 – Active Zones for STEP-01: Entering the Environment

### 6.2.2 NARRATIVE STEP 02: Loading Dataset.

The user expands the free-standing panel to view its notifications, revealing that one of their collaborative data sets has been modified and returned to them. The notification displays metadata such as project name, a description of the updated version, and a small 3D render of the space, along with the option to load the project or dismiss the notification.

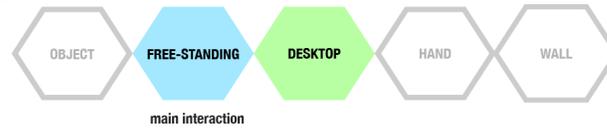

Figure 9 – Active Zones for STEP-02

### 6.2.3 NARRATIVE STEP 03: Working in the Space.

The user opens the project. The free-standing panel disappears, and the notifications are now on only the Wall panel. The object of focus appears in the space along with two Object zones that display hyper-specific information about the object's 3D relations and a dock to swap manipulation tools, a set of Free-standing panels that display data that, while closely related, does not necessarily reflect the 3D representation, and two zones on the Desk, one which mirrors the tool-swapping capability of the Object zone, and one that has more crude actions relating to the object and to the space in general. As the user swaps the tool on their pen, the pen's model should augment to reflect its current functionality.

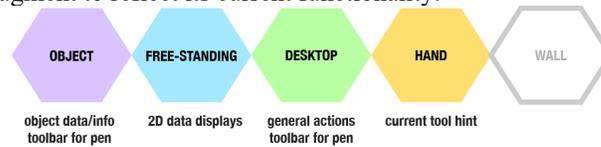

Figure 10 – Active Zones for STEP-03

### 6.2.4 NARRATIVE STEP 04: Interacting with Central 3D Representation.

The user picks up their pen and works closely with the 3D representation, interacting more with the 3D object and its Object zone, using the pen to take down precise point measurements relating to the object.

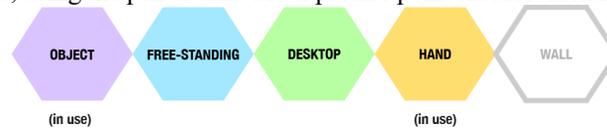

Figure 11 – Active Zones for STEP-04

### 6.2.5 NARRATIVE STEP 05: Finding Connections with 2D Representation.

Next, they lean back and look more specifically at the 2D data, interacting more with Free-standing panels upon which the data is presented. They notice that the data's text is a little difficult to read, so they modify the font size through the Desk settings menu before returning to the data panels.

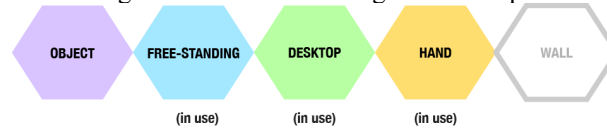

Figure 12 – Active Zones for STEP-05

### 6.2.6 NARRATIVE STEP 06: Exiting the Space.

Finally, the user has concluded their work. They look over everything in the space one last time, and they notice on a Wall panel that their collaborators sent them an email asking for an update. They move to save, export, and quit their environment using the Desk zone.

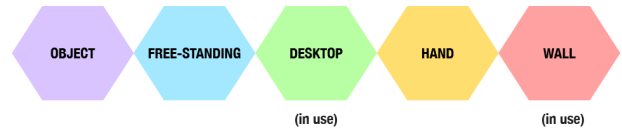

**Figure 13** – Active Zones for STEP-06

## 7 Conclusions

We propose a taxonomy for understanding the types of interactions needed in an MR data exploratory environment, and we present a design space that describes interaction zones to guide application development through natural, efficient and intuitive interaction. While our collaborators, scientist and clinicians, have found the our implementation of interactive spaces tailored for their needs quite engaging and useful, we believe that beyond our empirical experiences, our careful analysis of the general needs of our target users, coupled with supporting research into what considerations lead to better exploratory processes have lead us to the creation of a strong design space, which we are eager to see others apply.

**Acknowledgements**

We would like to thank Dr. Ava Wexler, MD from the School of Medicine at University of Rochester, who directly collaborated with us in the creating a deeper understanding of the mindset of clinicians as they navigate different data; Dr. Evan M. Zahn, MD at Cedars Sinai in taking us through the analysis phase of complex pediatric heart cases as well as exposing us to how data and data analysis informs the direct decisions being made during real operating room interventions, as well as all collaborators, patients, and NIH for trusting us with your data and spending time with us to have a deeper understanding of how your different scientific endeavors rely on visualization and direct data analysis in order to reach deeper knowledge, and collaborating with us in exploring how and when are MR technologies helpful, when they are hindrances, and where we can help in making your current processes any more efficient, effective, or reach a higher level of certainty.